\documentclass{article}

     \usepackage[final]{neurips_2019}

\usepackage[utf8]{inputenc} % allow utf-8 input
\usepackage[T1]{fontenc}    % use 8-bit T1 fonts
\usepackage{hyperref}       % hyperlinks
\usepackage{url}            % simple URL typesetting
\usepackage{booktabs}       % professional-quality tables
\usepackage{amsfonts}       % blackboard math symbols
\usepackage{nicefrac}       % compact symbols for 1/2, etc.
\usepackage{microtype}      % microtypography
\usepackage{amsmath}
\usepackage{graphicx}
\usepackage{subcaption}
\usepackage{url}
\setcitestyle{round}
\newcommand{\paragraphHdTop}[1] {\noindent\textbf{#1}}

\title{On Embeddings in Relational Databases} %\\ OR \\ How I Learned RDBMS is NOT a KG}
\author{%
  Siddhant Arora \quad Srikanta Bedathur \\
    CSE, IIT Delhi, \\
    Delhi, India. \\
    \texttt{\{Siddhant.Arora.cs515, srikanta\}@cse.iitd.ac.in} 
}

\begin{document}

\maketitle

\begin{abstract}
We address the problem of learning a distributed representation of entities in a relational database using a low-dimensional embedding. Low-dimensional embeddings aim to encapsulate a concise vector representation for an underlying dataset with minimum loss of information. Embeddings across entities in a relational database have been less explored due to the intricate data relations and representation complexity involved. Relational databases are an inter-weaved collection of relations that not only model relationships between entities but also record complex domain-specific quantitative and temporal attributes of data defining complex relationships among entities. Recent methods for learning an embedding constitute of a naive approach to consider complete \emph{denormalization} of the database by materializing the full join of all tables and representing as a knowledge graph. This popular approach has certain limitations as it fails to capture the inter-row relationships and additional semantics encoded in the relational databases. In this paper we demonstrate; a better methodology for learning representations by exploiting the underlying semantics of columns in a table while using the relation joins and the latent inter-row relationships. Empirical results over a real-world database with evaluations on similarity join and table completion tasks support our proposition.
  
%on the other hand, is used not only to model complex, heterogeneous relationships between entities but also stores specific (numerical) quantities (e.g., sales volume, salary, division-wise employee counts etc.), 
%   Distributed Representation via word embedding has gained immense popularity for various Natural Language Processing tasks since they capture useful syntactic and semantic properties of word based on its context. However, not a lot of research has been made to extend this idea to relational databases. We believe that this extension could solve complex analogy and similarity queries on databases. In this paper, we aim to learn representation for entities on relational databases to capture latent relation between database entities. We comprehensively evaluate the quality of these learned representation and empirically compare various methodologies for learning these embeddings. We further evaluate the usefulness of these embeddings on question answering on databases and provide evidence of its application on database completion. We also perform some preliminary investigation on how joining of tables improve quality of embeddings and believe our methodology could be adapted on star relational networks to find which joins could be avoided.
\end{abstract}

\section{Introduction}
Embedding based methods have recently gathered a lot of attention mainly due to their ability to generate a mapping of categorical/discrete variables to continuous vector spaces and thus their support towards neural network models. They have been applied to a variety of tasks namely, images \citep{alex}, text \citep{Word2vec}, knowledge graphs\citep{kg1}, time series \citep{vrnn} with limited work on effective embeddings for entities in databases with \emph{multiple} relational tables.
\cite{Ref1} introduced relational databases embeddings, but their proposed approach treated the row of a table as a sentence for learning word2vec embeddings of an entity and did not consider the problem involved in handling heterogeneous relations that exist between entities or inter-row relationships encoded in the databases. 

Operational databases consist of multiple tables with each table incorporating complex relations both within and across rows of tables. This complexity persists even
when we consider a fully \emph{denormalized} relational database, obtained by completing all the natural joins between primary-key–foreign-key columns. For instance, the real-world IMDB movie database (\cite{imdb}) contains among others, \texttt{Director (id, first\_name, last\_name)} and \texttt{Directors\_Genre (director\_id, genre, prob)}
tables which maintain information with each entity consisting of a \emph{director} with directed a movie and their genre preferences respectively. Specifically, the \texttt{Directors\_Genre.prob} is the observed probability of the corresponding director directing a movie of that genre. Naturally, this continuous-valued attributes cannot be captured accurately by a naive word2vec model operating directly on the table. Instead, by using the semantic knowledge of the data being modeled, we can generate a modified representation of the table to significantly boost the performance of the word2vec model.

Another possible view is to consider the table as a knowledge graph with a KG embedding method while learning representations. Apart from the limitations above, constructing a KG fails to capture inter-row relationships that are often implicit in the database and play an important role in the effective representation of entities. For instance, consider two tables \texttt{Movies (id, name, year, rank)} and \texttt{Movies\_Directors (director\_id, movie\_id)} connecting movie details with corresponding director(s). Now, two directors with similar trajectories in their directorial outcomes are expected to close in their representations. But neither of the above two models of learning representations of relational tables will be able to capture this effect. Instead, a specialized sequence embedding model such as an RNN with LSTM \citep{lstm} captures the temporal characteristics using the inter-row sequential relationships encoded in the database.

We further support our observations through empirically designing and evaluating alternate embedding strategies that utilize the semantics of columns in a relational table. Specifically, we consider using weighted sample views of the database table (which can be implemented in SQL efficiently) as the input to word2vec models for two strategies in the context of our movie database –viz., \emph{Director-Genre} sampling and \emph{Movie-rank} sampling. Later, we consider LSTM for capturing inter-record relations because of the temporal data encoded in the movie database. Naturally, other alternatives may need to be used in other forms of inter-record relationships are to be captured (e.g., spatial proximity of locations).

We consider our evaluations across two metrics; (i) Similarity queries, and (ii) data cell completion and demonstrate a novel coherent evaluation framework for similarity query task by using Milne-Witten \citep{witten} inlink score on Wikipedia as the baseline.

To summarise our contributions in this work:
\begin{enumerate}
\item We overcome the limitations associated with the previous methods and leverage semantic level information to effectively learn embeddings on databases.
\item Exploiting inter-row relation and time-series data that are highly common in database settings along with a rigorous evaluation across multiple baselines.
\end{enumerate}

\section{Related Work}
In this section we present a list of methods that are closely related to us namely, learning embedding on relation databases, learning global table embedding for table retrieval tasks and table augmentation.

\subsection{Learning embedding on relation databases}
\citet{Ref1} proposed the concept of learning embeddings on relational databases by using unsupervised, distributed representation to map database entities to latent space. Their approach to capture the semantic relationship between entities in a row in the database was aimed to solve complex \emph{cognitive} queries. \citet{RajeshDisclosure} further looked into novel challenges in this domain mainly addressing the degree of opacity of the disclosed database information.

\subsection{Learning global table embedding for Table Retrieval tasks}
Earlier approaches \citep{BalogTableSearch} have explored the use of global table embeddings for the task of returning related tables with the representation of table and query into multiple latent semantic space and then an  aggregated matching over all these spaces. \citet{RelatedTable} extend the above idea but instead of matching query with the table they matched tables with each other by representing both of them in latent semantic space.

\subsection{Table Augmentation}
Table augmentation was intended to fill missing entities in the table based on information from table corpus. \citet{InfoGather} looked into indirectly matching web table to input using Topic Sensitive Pagerank and augmentation framework. \citet{InfoGather+} extend the above idea by looking into matching continuous data streams like numeric data stream and time-series data by building a semantic graph. Recently focused have been shifted to other allied fields like row and column augmentation in \citet{EntiTables}. They focused only on those relational databases that have an entity focus. They looked into populating row with additional entries which has earlier been explored in (\citet{wang2015}, \citet{das2012}) as well as populating a column with additional fields for the entity. They came up with a learning methodology for both tasks that combined information from both knowledge graph and table corpus. Table augmentation was revisited in \citet{BalogAutoComplete} where they looked into probabilistic models that can be used to identify candidate value from data cell. Their approach mapped the target attribute to semantically similar column heading in table and predicate in Knowledge Base. Finally, they came up with novel Learning to Rank-based approach to rank these candidate cells.

\section{Baseline Methodology}

\subsection{Simple Word2Vec on the Denormalized Table}
Following \citet{Ref1} we learn word2vec (\citet{Word2vec}) for entity representation with each row in the database represented as sentence and database entities as words. Then we include simple word2vec model among our baselines.

\subsection{Knowledge Graph Constructed on the Denormalised Table}
Here we demonstrate the procedure of creating a knowledge graph from relational databases. For each row in the database, there are multiple column entities. We create a relation between every pair of columns; for example, between column i and j, we create the relation "i\_j". Then for every row in the database, each pair of entities are joined by a relation in the knowledge graph, for example, if entity a occur in column "i1" and entity b occur in column "j1" in the same row in the database, then a and b get \emph{joined} by relation "i1\_j1" directed from a to b and \emph{joined} by relation "j1\_i1" directed from b to a.

\begin{figure}[h!]
  \centering
    \includegraphics[width=\linewidth,height=4cm,keepaspectratio]{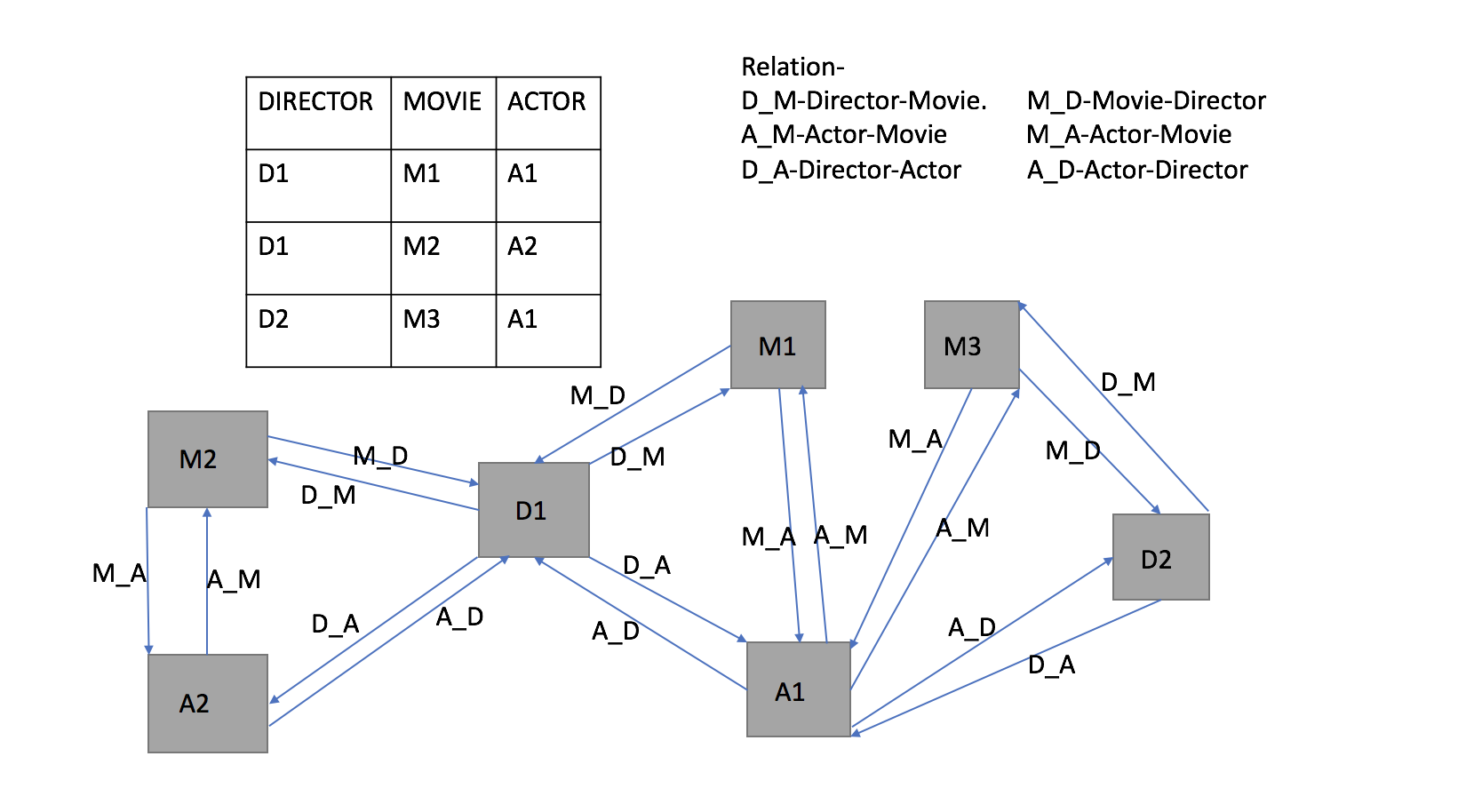}
 \caption{Creation of Knowledge Graph from a Relational Table}
  \label{fig:TransH}
\end{figure}

We then use TransH \citep{TransH} to learn embeddings for entities in the knowledge graph. TransH can learn one-to-N relation which supports its suitability in the setting of a relational database.

\section{Column Semantics for Denormalised Table Sampling}
Databases are complicated entities and they usually consist of continuous stream of data (like integers) whereas word2vec has been traditionally applied to discrete data. In our setting, we had \texttt{movie-rank} and  \texttt{director-genre} as a continuous stream of rational numbers between 0-10 and 0-1 respectively. To convert them into discrete distribution, we restrict movie rank to closest integer and director genre to the closest multiple of \emph{$0.1$}. However, we still would not be able to capture the inherent sequential properties in continuous streams of data. It is also important to note that in a sentence, words that are closer to each other in the sentence are more similar to each other. This approach faces a problem when we look at a row in a database since every entity in the row has \emph{many-to-one} relation with every other entities in a row. As a result, each word’s context should span the entire row. In our experiments, we exhaustively criticize this approach when we look into the impact of changing the context window.

\subsection{Genre Sampling}
In our dataset, each director has multiple genres with associated director genre probability. Here it is important to note that the genre probabilities are not only continuous streams of data but also exhibit sequence and order properties. To motivate the above statement, the example where Director A has genre mystery with probability "0.6", Director B has genre mystery with probability "0.2" and Director C has genre mystery with probability "0.7".
Thus, it appears that Director A should be considered to be more similar to Director C than Director B. However, it is unfair to expect from the word2vec model to learn that 0.7 and 0.6 are closer values and hence A and B are more closely related to each other. Thus, we come up with a simple fix to this problem to introduce this semantic ordering into our embeddings. We use this genre probability to sample genres for a director. We expect to make the model learn that genre associated with the director with higher probability is more closely related and hence should be sampled more times in the director's context.

We explain our methodology with an example. Suppose Director has 3 genres \emph{A, B and C} with probabilities $p_A,p_B$ and $p_C$. Then Genre \emph{A} will be sampled with probability $\frac{p_A}{p_A + p_ B + p_C}$.  Genres are sampled for each director 6 times according to the above-mentioned probability. We also shuffle the entire sentence so that we can avoid changes results by a different order of joining of tables.

\begin{figure}[h!]
  \centering
    \includegraphics[width=\linewidth,height=5cm,,keepaspectratio]{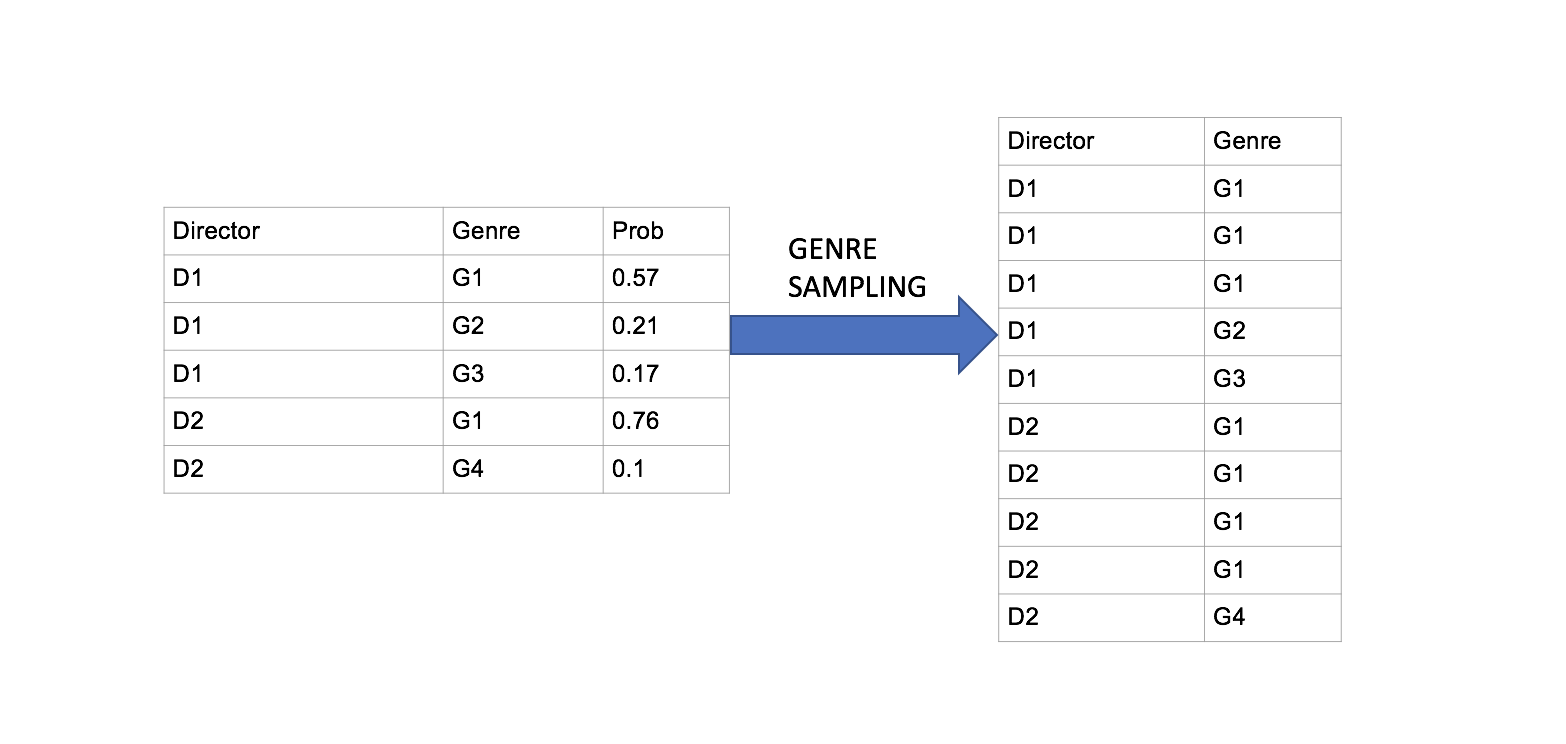}
 \caption{Analyzing Column Semantics}
  \label{fig:Genre}
\end{figure}

\subsection{Movie Rank Sampling}
In our data set each director has movie rank associated with each directed movie with each movie having a rank associated with it. Here for each director movies get a sample based on the movie rank. The motivation behind this variant is the fact that a director is more known for his popular films. This variant is built on top of genre sampling.

\section{Incorporating inter-record Relationships}
\begin{figure}[h!]
  \centering
   \includegraphics[width=0.8\linewidth,height=5cm,keepaspectratio]{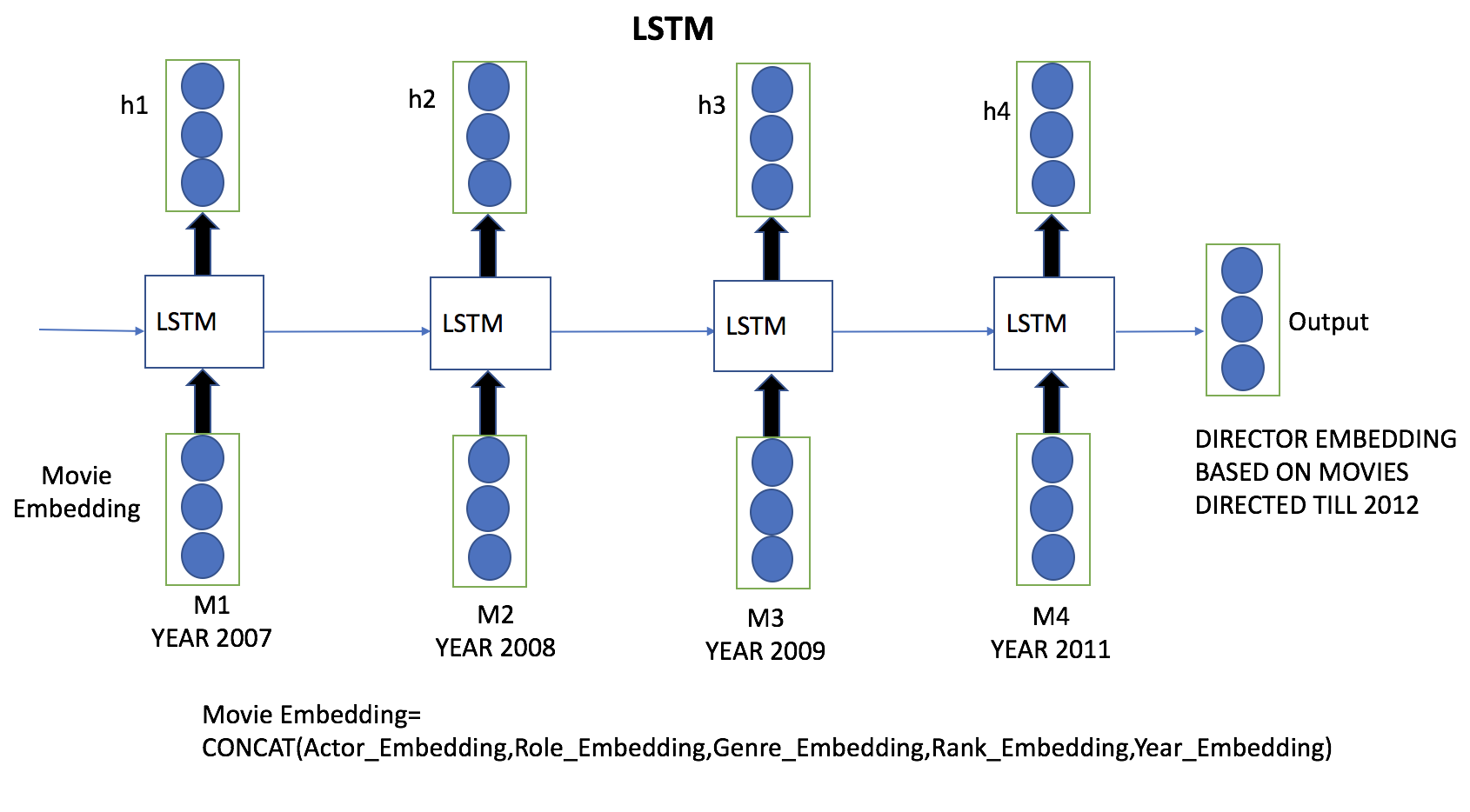}
 \caption{LSTM for Director Embeddings}
  \label{fig:LSTM}
\end{figure}
We can represent a time series for each director consisting of the movies directed by him sorted temporally. As discussed before, most of the embedding frameworks cannot effectively learn embeddings on time series data since they can not capture sequentially and thus cannot retrieve the relation that the year 1976 comes before 1999 which comes before 2002. This is important because directors who direct movies during a similar period should be similar. Moreover, the director’s interest may change over time and hence we could use time series to build time-sensitive embedding for the director. But the word2vec model is incapable to learn the differences associated with these drastic changes in temporal characteristics.
Thus we use time series for each director based on their movies to create "time-sensitive" director embedding based on his movie preference. We give all the movies directed by the director till time t as input time series. All the parameters of the movie at time t’ are provided as input at time t’ to LSTM. For our relational database, the parameters for each movie are actor, role,movie\_genre, movie\_year, and movie\_rank. Since a movie has multiple values for parameter actor, role and genre, we randomly select one of these values for these parameters. Embedding of the movie is generated by concatenating actor, role,movie\_genre, movie\_year, and movie\_rank embedding. Movie embedding at time t’ is provided as input at time t’ to LSTM. The LSTM output is then passed through a fully connected layer to bring it to the same dimension as movie embedding. Cosine similarity
between this generated director embedding and movie embedding is used to predict the movie directed by the director at time t+1. We call the generated director embedding as director embedding based on movie preference.

We then optimize the director embedding by maximizing its cosine similarity with the movie directed by him at time t+1.To do so, we randomly sample 5 movies not directed by the director. We take the dot product of this director embedding with movie embedding of correct last movie as well as movie embedding of 5 randomly negative sampled. We then take softmax over this output obtained after dot product and optimize Negative Log-Likelihood loss to train the LSTM.

\subsection{LSTM variants}
\paragraphHdTop{Joint Loss variant}
It was observed that learned director embedding was close only to the movie embedding at time t+1 and has lost the information of all other movies directed by the director. Hence the Loss function was computed for each of the correct movies (i.e. movies used as input to LSTM as well as the movie at time t+1) and net loss was computed as the sum of all these losses. All other LSTM variants are built on top of Joint Loss Variants.\\
\paragraphHdTop{Actor sampling}
We identified that not a lot of information can be captured by only one random actor since most of the movies tend to have a large number of actors. Thus we come up with a variant where we randomly sample 3 actors and 3 roles for each movie. We then combine actor embedding for these actors based on 2 operations, one is simple aggregation by averaging whereas the other was concatenating them and passing them to linear layer to get movie’s aggregated actor embedding. A similar operation was performed to generate the movie’s aggregated role embedding. We compare the
performance of these 2 approaches in the Results section.\\
\paragraphHdTop{Popular actor variant}
We first identify the popularity of an actor based on the number of movies he has worked on. We then choose the top 3 most popular actors for each movie and their corresponding roles in those movies to create movie embedding.
\section{Evaluation}
\subsection{Data}
For our experiments, we use the IMDB dataset which consists of 7 tables which when combined comprise a total of 21 columns. A table describes subtopics like director, director genre, movie, actor, etc. with columns referring to entities in this subtopic like director id, director first name, director last name for the director table.

% \begin{figure}[h!]
%   \centering
%     \includegraphics[width=\linewidth]{Presentation_Folder/Data.png}
%  \caption{Data used for Experimentation. Picture from https://relational.fit.cvut.cz/dataset/IMDb}
%   \label{fig:Data}
% \end{figure}
\subsection{Models}
All word2vec models were trained on the entire dataset with embedding dimension 300 for 10 epochs. Just to see the effect of context window size, we experimented with a context window of size 5 for basic word2vec [Base w2v(Context 5)] whereas, for other variants, context window was taken as the size of a sentence formed from each row in the table. We also looked into the effect of the relative position of the columns. We brought the movie column next to the director column and then trained word2vec (Base w2v
(Close)) just to see if there are significant changes induced by the relative position of columns. 
TransH models were run for 1000 iteration at a learning rate 0.001 with embedding dimension 50 for each entity and relation.
For LSTM implementation, the movies directed by the director are sorted based on year. Now we pick 6 movies directed by the director- the first 5 movies are selected for the training set and the last movie for the test set. We also randomly sample 5 movies not directed by the director as negative samples for training. For each director, 80\% of movies go into the training set, 10\% into the validation set and the remaining 10\% into the test set.
Now in these 80\% movies in the training set, We randomly sample 5 movies from the director’s training set and 5 movies not directed by the director as negative samples for training. We repeat steps described before (len(movies in the director’s training set)/5) number of times for each director. Then first 4 movies are used as input to LSTM whereas movie 5 is used as a movie at t+1 to compute loss for LSTM. The model is trained using Adam Optimiser with a learning rate of 0.001 with a batch size of 1024. All the other variants of LSTM are also trained in similar training methodology.
\subsection{Evaluation using WikiLinks}
Now, the task before our hand was to find a way to evaluate our approach. We take inspiration from methodologies to obtain semantic relatedness between Wikipedia Entities. Milne Witten Inlink Score (\citet{MilneScore}) uses the hyperlink structure of Wikipedia to measure this semantic relatedness. To evaluate our approach, we looked at the top 63 popular directors according to IMDB. We create a list of 100 most similar directors based on Milne Witten Inlink score as gold sample.We evaluate our models by generating a list of 100 most similar directors based on similarity model and comparing it with the gold sample in terms of NDCG and Precision. For NDCG computations, score 5 was given to top20 results, score 4 was given to top20-top40 results, score 3 was given to top40-top60 results, score 2 was given to top60-80 results and score 1 was given to top80-100 results and 0 to others. For precision, we assumed the top 100 most similar directors in our gold sample as relevant directors to the given director.\\
\subsection{Evaluating Database Completion}
\begin{figure}[h!]
  \centering
    \includegraphics[width=\linewidth,height=4cm,keepaspectratio,]{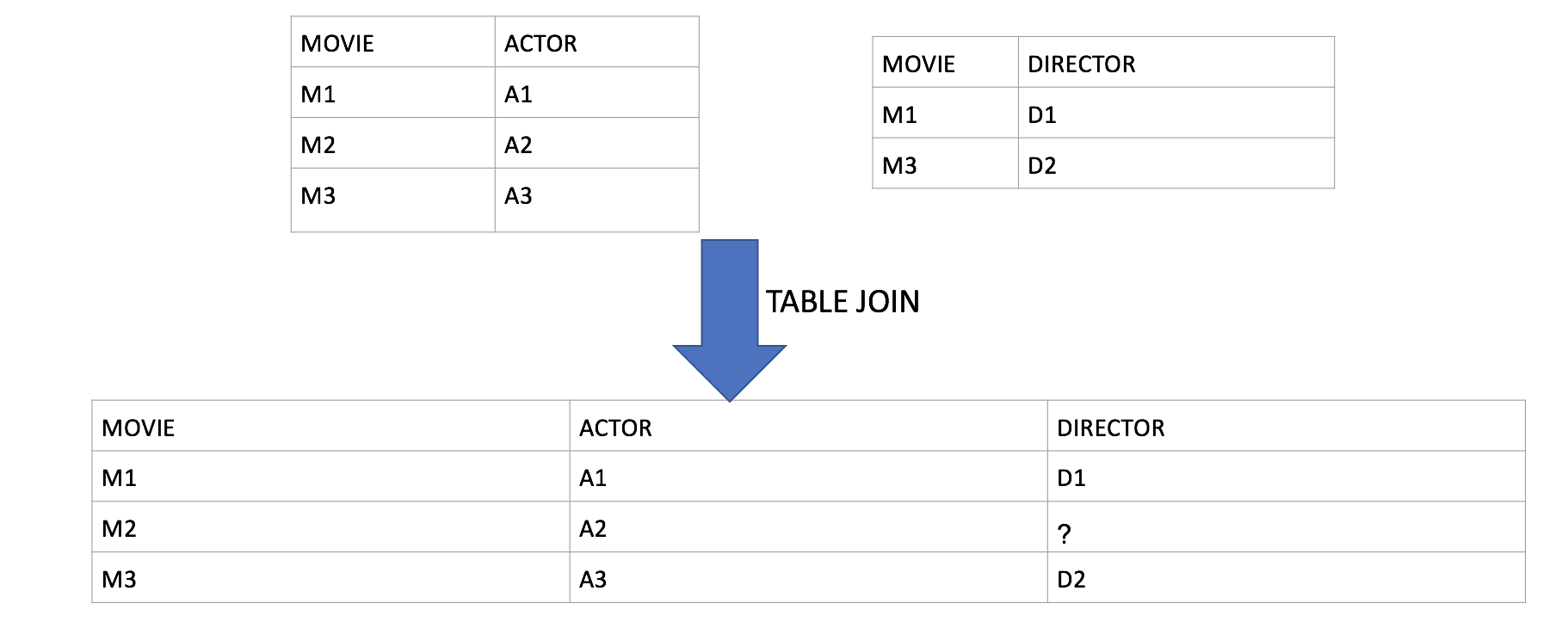}
 \caption{Envisioned Auto Completion of Data Cells in table}
  \label{fig:DataComplete}
\end{figure}
We randomly select 20\% of movies directed by the director for the test set and put remaining movies in train set for word2vec variants. Then for every movie in the test set, We randomly sample 99 directors as negatively sampled directors. Finally, for each movie in the test set, We compute its similarity with correct as well as negatively sampled directors. We then rerank this list of candidate directors based on similarity scores and compute precision values on this reranked list.\\

We also performed time-sensitive Database completion where instead of randomly selecting 20\% of movies, we select the last 20\% of movies in the test set. Then for each movie in the test set, we utilized the following methodology. If the movie was directed at time t, we use movies directed by director before time t but not more than 20 years before time t to create "time-sensitive" director embedding. We use these time-sensitive director embedding for the correct director as well as 99 negatively sampled directors and followed methodology similar to one in the paragraph before. For this, we also selected 20\% movies in the validation set and the remaining 60\% movies for the train set. The embeddings used in LSTM
were initialized by embeddings learned from genre sampling models and were not backpropagated in our training.

\section{Results}
\subsection{Similarity Queries}
\label{tab:WikiLinks}
\begin{table}

\centering
 \begin{tabular}{@{}l l l l l @{}} 
 \toprule
 Model & Prec@10 & Prec@20 & NDCG@10 & NDCG@20 \\
 \midrule
 Base w2v (Context 5) & 0.158 {\tiny (T)} & 0.127 & 0.124 {\tiny (T)} & 0.104 {\tiny (T)}\\ 
 Base w2v (Close) & 0.145 {\tiny (T)} & 0.132 & 0.114 {\tiny (T)} & 0.102 {\tiny (T)} \\
 Base w2v & 0.169 {\tiny (T)} & 0.142 {\tiny (T)} & 0.126 {\tiny (T)} & 0.111 {\tiny (T)}\\
 Genre w2v & 0.196 {\tiny (TsC)} & 0.161 {\tiny (TSC)} & 0.144 {\tiny (TC)} & 0.124 {\tiny (Tc)}\\
 Movierank w2v & 0.176 {\tiny (T)} & 0.149 {\tiny (T)} & 0.136 {\tiny (T)} & 0.120 {\tiny (T)}  \\
 TransH & 0.106 & 0.112 & 0.079 & 0.078 \\
 LSTM loss & 0.155 {\tiny (T)} & 0.140 & 0.120 {\tiny (T)} & 0.110 {\tiny (T)}\\
 \bottomrule
\end{tabular}
\caption{Results on the Similarity Queries. Significance tests were conducted using a two-tailed paired Student's t-test. Uppercase or lowercase characters in brackets indicate statistical significance with $p<0.05$ or $p<0.10$, respectively, over the Base w2v (B/b), Base w2v (Context 5) (S/s),Base w2v (Close) (C/c), Genre w2v (G/g), Movierank w2v (M/m), TransH (T/t) and LSTM Joint Loss Variant(L/l).}
\end{table}
Results for solving similarity queries on databases by using wikilinks as golden standard are shown in table 1. We can see that Word2vec variants seem to be the best performing models. Genre sampling model outperforms all models in terms of NDCG and word2vec followed by MovieRank sampling models. We also see that increasing Context Window Size to the entire length of the sentence seems to improve performance. This is because in the database the columns that are adjacent to each other may not necessarily be similar to each other. Thus in a database, entities are affected by all adjoining entities in a row in the database. Also, the relative position of these entities does not affect performance as there is no statistically significant difference between them. Hence by keeping the context window large enough we can overcome the effect of the relative position of columns. Hence we just shuffle row before performing word2vec in Genre Sampling and Movie Rank Sampling models. We can conclude from this table that sampling based on integer values may not always produce the most idealistic results. In this case, movies that might not be popular can still give crucial insight that could be used as a distinctive feature of a director like what genre of movies are preferred by the director and what actors do he usually likes to work with.

It was also observed that about half of the directors in our database were not Wikipedia entities. This might adversely taint our results since there is a possibility that there exists a director that is very similar to our candidate director but it does not happen to be a Wikipedia entity. Thus we recompute the result for similarity queries but this time reranking only those directors that are in the Wikipedia database. The results can be seen in table 2. The trends are similar to as seen before but the Precision and NDCG values are much better however we can see that Genre Sampling and Movie Rank Sampling variants of word2vec are statistically significantly better than word2vec and TransH baseline models.

\label{tab:WikiLinksNew}
\begin{table}

\centering
 \begin{tabular}{@{}l l l l l @{}} 
 \toprule
 Model & Prec@10 & Prec@20 & NDCG@10 & NDCG@20 \\
 \midrule
 Base w2v (Context 5) & 0.182 & 0.155 & 0.142 & 0.123\\ 
 Base w2v (Close) & 0.153 & 0.139 & 0.122 & 0.108 \\
 Base w2v & 0.188 & 0.161 & 0.142 & 0.127\\
 Genre w2v & 0.248 {\tiny (TSBC)} & 0.208 {\tiny (TSBC)} & 0.186 {\tiny (TSBC)} & 0.162 {\tiny (TSBC)}\\
 Movierank w2v & 0.235 {\tiny (TSBC)} & 0.2 {\tiny (TSBC)} & 0.187 {\tiny (TSBC)} & 0.164 {\tiny (TSBC)}  \\
 TransH & 0.161 & 0.15 & 0.115 & 0.106\\
 LSTM loss & 0.192 & 0.165 & 0.146 & 0.132\\ 
 \bottomrule
\end{tabular}
\caption{Results on the Similarity Queries for reranking only those entities in Wikipedia. Significant test performed the same as before.}
\end{table}
% We also tried to evaluate if there are any difference in result over performance of these semantic queries if we rerank only popular directors i.e. those that have directed atleast 5 movies. We can see the results in table 3. We see that here MovieRank Sampling performs better than genre sampling. 
We also tried to evaluate if there is any difference in result over the performance of these semantic queries if we re-rank only popular directors i.e. those that have directed at least 5 movies. We can see the results in table 3. We see that here MovieRank Sampling performs better than genre sampling. We attribute this to the reasoning that for popular directors, the similarity between popular movies serve as an important factor for semantic similarity. Other trends are similar to as seen above.

\label{tab:WikiLinks1}
\begin{table}

\centering
 \begin{tabular}{@{}l l l l l @{}} 
 \toprule
 Model & Prec@10 & Prec@20 & NDCG@10 & NDCG@20 \\
 \midrule
 Base w2v  & 0.202 {\tiny (T)} & 0.175 & 0.150 {\tiny (T)} & 0.134 {\tiny (T)}\\
 Genre w2v & 0.229 {\tiny (T)} & 0.210 {\tiny (TlB)} & 0.174 {\tiny (T)} & 0.165 {\tiny (TB)}\\
 Movierank w2v & 0.242 {\tiny (TB)} & 0.218 {\tiny (TLB)} & 0.181 {\tiny (Tb)} & 0.164 {\tiny (TB)}\\
 TransH & 0.145 & 0.154 & 0.101 & 0.103\\
 LSTM loss & 0.194  & 0.173  & 0.153 {\tiny (t)} & 0.137 \\ 
 \bottomrule
\end{tabular}
\caption{Results on the Similarity Queries for reranking only those directors that have directed at least 5 movies. A significant test performed the same as before.}
\end{table}

\subsection{Database Completion}
\begin{figure}[h!]
  \centering
  \begin{subfigure}[b]{0.4\linewidth }
    \includegraphics[width=\linewidth,height=4cm,keepaspectratio,]{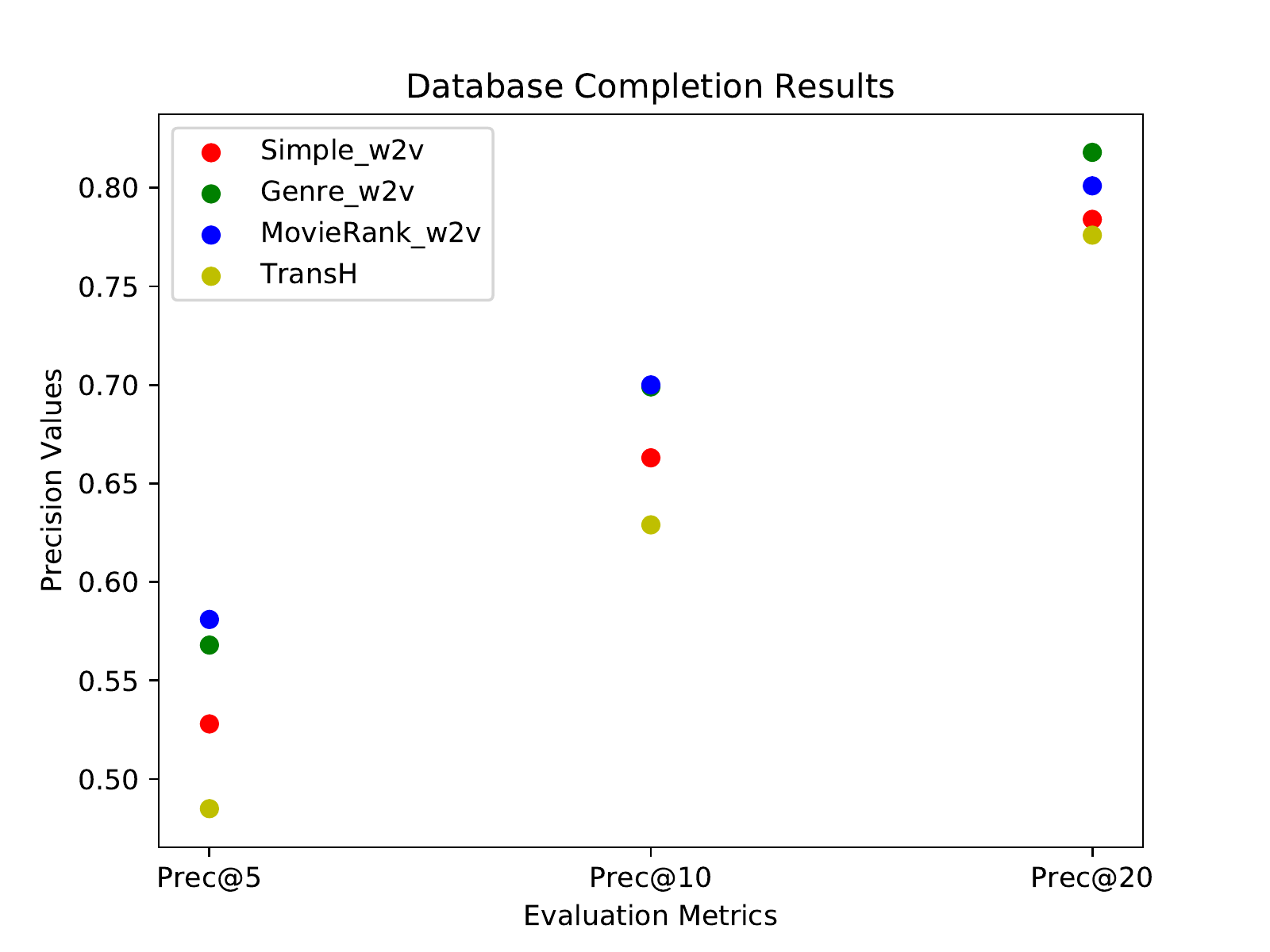}
    \caption{Database Completion Results Random Removal}
  \end{subfigure}
  \begin{subfigure}[b]{0.4\linewidth}
    \includegraphics[width=\linewidth,height=4cm,keepaspectratio,]{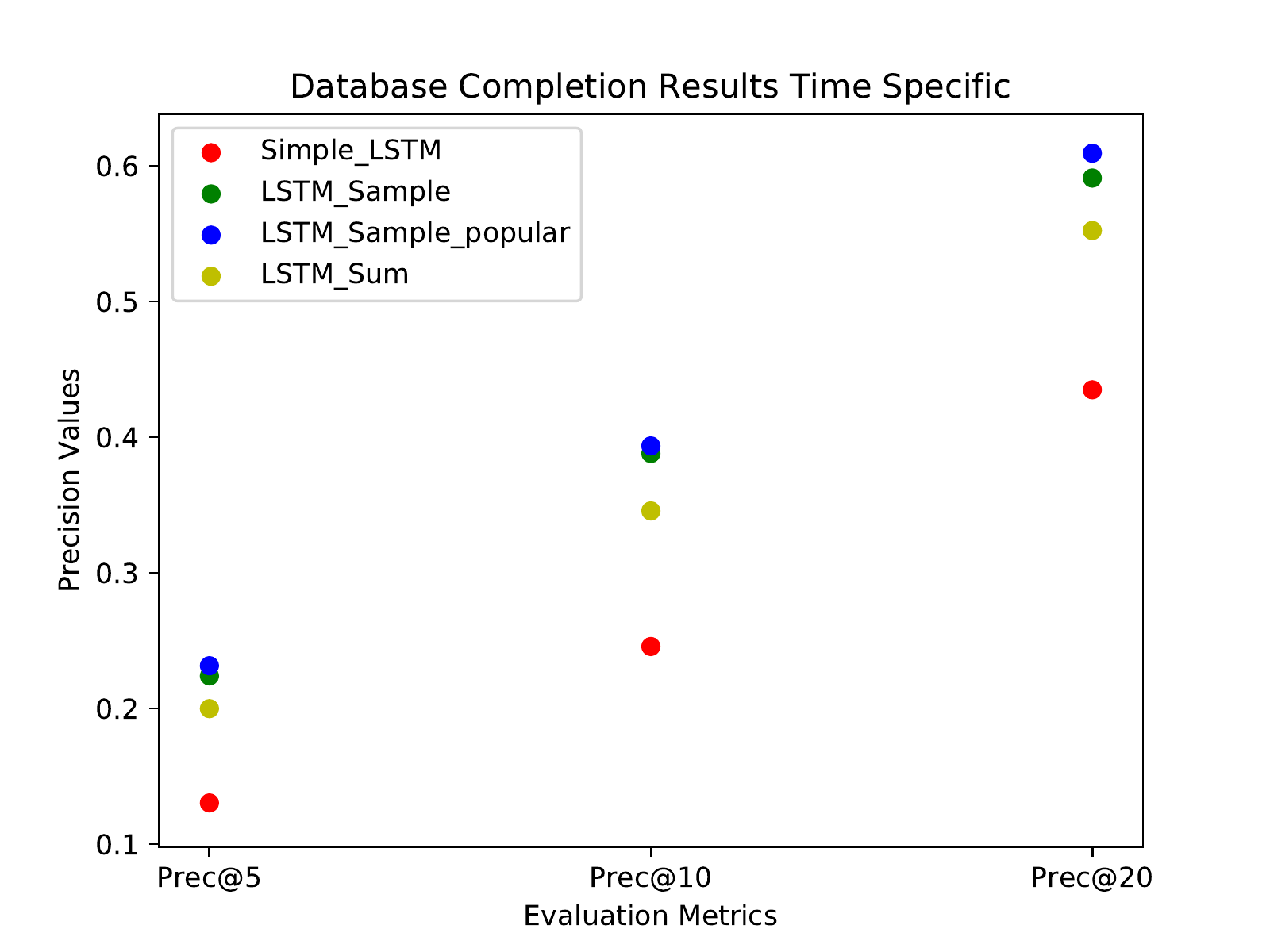}
    \caption{Database Completion Results Time Specific}
  \end{subfigure}
  \caption{DataBase Completion Results}
  \label{fig:Database}
\end{figure}
Now we look into results of Database Completion for word2vec variants in 5. We see that again Genre Sampling and MovieRank Sampling outperform the basic Word2Vec model. Moreover, we see a very positive performance of nearly 0.8 for precision@20 providing evidence that our approach can be successfully utilized in the auto-completion of data cells. In the second graph, we do time-specific database completion. We see that LSTM variants perform better than simple Joint Loss LSTM. LSTM Sampling and LSTM popular sampling variant model seems to be best performing models with both achieving around 0.6 Precision@20.
% \begin{table}
% \caption{DataBase Completion Results}
% \label{tab:DatabaseComplete}
% \centering
% %\setlength{\tabcolsep}{8pt}
%  \begin{tabular}{@{}l r r r @{}} 
%  \toprule
%  Model & Prec@1 & Avg Test Rank \\
%  \midrule
%  Simple w2v & 0.248 & 13.7\\
%  Genre w2v & 0.285 & 11.7 \\
%  MovieRank w2v & 0.301 & 12.9 \\
% %  LSTM &  0.2776 & 33.3 \\
% %  LSTM Actor & 0.260 & 21.0 \\
% %  LSTM loss & 0.204 & 13.3\\ 
%  \bottomrule
% \end{tabular}

% \end{table}

% \paragraphHdTop{MovieQA\\}
% \\
% \begin{figure}[h!]
%   \centering
%   \begin{subfigure}[b]{0.4\linewidth}
%     \includegraphics[width=\linewidth]{MovieQA1.pdf}
%     \caption{MovieQA Results based on Movie Index}
%   \end{subfigure}
%   \begin{subfigure}[b]{0.4\linewidth}
%     \includegraphics[width=\linewidth]{MovieQA2.pdf}
%     \caption{MovieQA Results based on Movie Name}
%   \end{subfigure}
%   \caption{MovieQA Results}
%   \label{fig:MovieQA}
% \end{figure}
% We see that the results of Question Answering on Dataset are very encouraging as can be seen in \ref{fig:MovieQA}. We see that when performing Question Answering on datasets, word2vec variants heavily outperform the simple word2vec model. We also observe that when we have tyo use only Movie Name, the performance of these word2vec variants detoriate significantly. This can be attributed to the fact that entities in movie name can be general and it is difficult to relate movie by its movie name to particular genre. However, still word2vec variants outperform base model for Question Answering based on Movie Name.

\section{Conclusions}
Learning entity embeddings in a relational database presents interesting challenges over traditional models over text data or knowledge graphs due to various complex data semantics that co-exist. We believe that our results, though preliminary, demonstrate not only the need for embedding entities in relational databases, but also to guide these embeddings carefully by using data semantics. Going further, we plan to investigate how to formally share these data semantics declaratively with representation learning frameworks over relational databases. 

% \section*{References}

% References follow the acknowledgments. Use unnumbered first-level heading for
% the references. Any choice of citation style is acceptable as long as you are
% consistent. It is permissible to reduce the font size to \verb+small+ (9 point)
% when listing the references. {\bf Remember that you can use more than eight
%   pages as long as the additional pages contain \emph{only} cited references.}
% \medskip
\bibliographystyle{plainnat}
\bibliography{main}
\small

% [1] Alexander, J.A.\ \& Mozer, M.C.\ (1995) Template-based algorithms for
% connectionist rule extraction. In G.\ Tesauro, D.S.\ Touretzky and T.K.\ Leen
% (eds.), {\it Advances in Neural Information Processing Systems 7},
% pp.\ 609--616. Cambridge, MA: MIT Press.

% [2] Bower, J.M.\ \& Beeman, D.\ (1995) {\it The Book of GENESIS: Exploring
%   Realistic Neural Models with the GEneral NEural SImulation System.}  New York:
% TELOS/Springer--Verlag.

% [3] Hasselmo, M.E., Schnell, E.\ \& Barkai, E.\ (1995) Dynamics of learning and
% recall at excitatory recurrent synapses and cholinergic modulation in rat
% hippocampal region CA3. {\it Journal of Neuroscience} {\bf 15}(7):5249-5262.

\end{document}